# Optical absorption in graphene integrated on silicon waveguides

Huan Li[1], Yoska Anugrah[1], Steven J. Koester[1*], and Mo Li[1†]

[1]*Department of Electrical and Computer Engineering, University of Minnesota, Minneapolis, MN 55455, USA*

ABSTRACT

**To fully utilize graphene's remarkable optical properties for optoelectronic applications, it needs to be integrated in planar photonic systems. Here, we demonstrate integration of graphene on silicon photonic circuits and precise measurement of the optical absorption coefficient in a graphene/waveguide hybrid structure. A method based on Mach-Zehnder interferometry is employed to achieve high measurement precision and consistency, yielding a maximal value of absorption coefficient of 0.2 dB/µm when graphene is located directly on top of the waveguide. The results agree with theoretical model utilizing the universal ac conductivity in graphene. Our work provides an important guide for the design and optimization of integrated graphene optoelectronic devices.**

Graphene has many remarkable optoelectronic properties that are highly desirable for photonic applications but unobtainable or non-optimal in other CMOS-compatible material systems[1,2]. These include ultra-wideband absorption[3,4], controllable inter-band transition[5,6], saturable nonlinear absorption[7] and high-mobility ambipolar carrier transport[8,9]. Furthermore, the ability to produce high-quality graphene at the wafer scale[10], transfer and integrate graphene onto heterogeneous substrates[11], and pattern it into planar devices[12,13], makes practical application of graphene for optoelectronic devices imminent. Therefore, incorporating graphene with Si-compatible photonics is a very promising approach toward the realization of practical high-speed, low-power photonic systems integrated with CMOS circuits[14]. Many graphene based devices have been demonstrated recently, including photodetectors[15,16], modulators[17,18], polarization controllers[19] and ultra-fast pulsed lasers[20,21].

---

* Corresponding author: skoester@umn.edu

† Corresponding author: moli@umn.edu

The fundamental optical properties of graphene can be most conveniently investigated with free-space optics using a normal incidence configuration[22-25]. However, because graphene is a pure two-dimensional material, approaches to integrating graphene-based optoelectronic devices with CMOS-compatible materials (generally in a planar waveguide configuration) are needed in order to fully realize graphene's potential as the foundational material of an optoelectronic device technology. For example, the distinctive conical and gapless electronic band structure of graphene leads to a universal value for the interband-transition-induced ac conductance of $e^2/4\hbar$ over a wide spectral range from visible to infrared[3,4]. For normal incidence onto single layer graphene (SLG), this conductance converts to a constant absorption coefficient of $e^2/4\varepsilon_0\hbar c = 2.3\%$ [23]. This is a remarkably-high value considering graphene is only one atomic layer thick. In such a normal incidence configuration as shown in Fig. 1a, however, the interaction length between the optical field and the graphene is limited to the thickness of graphene (~1.1 nm) — the result is that 97.7% of the optical power is transmitted without absorption. If graphene is utilized as a wideband photodetector, this interaction length limitation leads to very low external quantum efficiency (and therefore responsivity) even if the internal quantum efficiency in the graphene can be high[15]. Similarly, to exploit other optical properties of graphene, such as electro-absorption modulation, saturable absorption, and polarization conversion, a long interaction length is necessary, and thus, normal incidence is not suited.

The strategy to achieve a longer light/graphene interaction length is to adopt a coplanar configuration by integrating graphene on a waveguide or fiber, so that the interaction length is decoupled from the graphene's thickness, and only limited by the length of the device. In such a graphene/waveguide hybrid the graphene resides in the evanescent field of the guided optical mode of the waveguide or fiber, leading to a much lower unit-length absorption coefficient compared with that of the normal incidence configuration. The obtainable interaction length, however, can be many orders of magnitude longer than using normal incidence, and can be made long enough to achieve nearly complete optical absorption as illustrated in Fig. 1b. Recently, such a configuration has been implemented to achieve a fiber-based polarization controller[19] and a waveguide-integrated electro-absorption modulator[17]. To further optimize and evaluate the performance capability of these graphene/waveguide hybrid devices, detailed understanding of the light/graphene interaction in the waveguide configuration is necessary, but has not been reported. In this paper, we demonstrate integration of graphene in a silicon photonic circuit and precisely determine the linear absorption coefficient and its dependence on the graphene/waveguide separation. To perform these measurements, we have adopted an interferometric method, which provides the necessary precision to confirm that the fundamental nature of the light/graphene interaction is the same in the waveguide integration as that revealed with normal incidence configuration. These results provide an important guide for the future design and optimization of integrated graphene optoelectronic devices.

Using Poynting's theorem of energy conservation in a dissipative media[26], the absorbed optical power in graphene can be understood as the resistive dissipation (or Ohmic loss) therein, which can be expressed by

$$\int_S \langle Q \rangle_s \, d\mathbf{r}^2 = \frac{1}{2} \int_S \mathbf{J}_s \cdot \mathbf{E}_t \, d\mathbf{r}^2 = \frac{\sigma_0}{2} \int_S |\mathbf{E}_t|^2 \, d\mathbf{r}^2 \qquad (1)$$

where $\sigma_0 = e^2/4\hbar = 6.08 \times 10^{-5}\, \Omega^{-1}$ is the ac conductance of graphene for interband transition, $\mathbf{E}_t$ is the in-plane component of the transverse electric field, $\mathbf{J}_s$ is the induced surface current. $\langle Q \rangle_s$ is the time-averaged resistive dissipation per unit area and the integral is performed over the surface of graphene $S$. Application of this relation to normal incidence with input optical intensity of $I_{inc} = \varepsilon_0 c |\mathbf{E}_t|^2/2$ gives the absorption coefficient of $\langle Q \rangle_s / I_{inc} = \sigma_0/\varepsilon_0 c = 2.3\%$, the same as that derived from the Fresnel formulas[3]. In the graphene-on-waveguide configuration, the dissipation $\langle Q \rangle_s$ leads to a linear absorption coefficient as given by:

$$\alpha = -\frac{1}{P(z)} \frac{dP(z)}{dz} = \frac{1}{P(z)} \int_L \langle Q_s(x) \rangle \, dl = \frac{\sigma_0}{2 P(z)} \int_L |E_t(x, y_0)|^2 \, dx \qquad (2)$$

where $x$ is the transverse direction, $z$ is the propagation direction of the mode (see Supporting Information for details). The graphene layer has a width of $L$ and is at a distance $y_0$ above the top surface of the waveguide with a cladding layer in-between. For the quasi-TE mode of the waveguide, the transverse electrical field $\mathbf{E}_t$ in the graphene consists predominantly of the $E_x$ component which decays exponentially as a function of distance $y$ outside the waveguide. Fig. 1c shows the finite-element simulation result of the quasi-TE mode of a typical silicon waveguide and the profile of the resistive dissipation distribution $\langle Q_s(x) \rangle$ in the graphene layer. Because the graphene resides in the evanescent field of the waveguide mode, the absorption coefficient decreases exponentially with $y_0$ as $\alpha(y_0) = \alpha_0 e^{-2\gamma y_0}$, where $\gamma$ is the field decay constant outside the waveguide.

To accurately determine the absorption coefficient in the graphene/waveguide hybrid, we integrate graphene on a Mach-Zehnder interferometer (MZI) circuit (Fig. 2a). The two arms of the interferometer can be covered by different lengths of graphene. Compared with conventional methods which determine the absorption coefficient by measuring the ratio between the input and the output optical power, the interferometric method employed here determines the ratio of power loss between a sensing interferometer arm and a reference arm from the extinction ratio (ER) of the interference fringes in the transmission spectrum. Thus, the large uncertainty in determining the absolute power levels and the optical coupling efficiency in the conventional method can be avoided. The output power of a MZI can be expressed by

$$P_o = (P_{in}/4)\left[e^{-\alpha l_1} + e^{-\alpha l_2} + 2e^{-\alpha(l_1+l_2)/2}\cos\left(2\pi n_{eff}\Delta L/\lambda\right)\right] \tag{3}$$

Here $\alpha$ is the linear absorption coefficient in the graphene/waveguide hybrid. The propagation loss in a silicon waveguide without graphene has a typical value of 3 dB/cm[27] and can be neglected when compared with $\alpha$ of the graphene/waveguide hybrid. Thus, only the lengths ($l_1$ and $l_2$) of the hybrid section in each interferometer arm (Fig. 2a) are of concern. The presence of the graphene layer induces negligible change in the real part of the waveguide effective index, so the phase difference between the two arms of the MZI is determined by $n_{eff}$, the effective index of the waveguide mode, $\Delta L$, the path length difference between the two arms and $\lambda$, the laser wavelength. As the input wavelength is swept, interference fringes are measured in the output optical power spectrum with an extinction ratio given by

$$\text{ER} = \frac{P_{max}}{P_{min}} = \frac{e^{-\alpha l_1} + e^{-\alpha l_2} + 2e^{-\alpha(l_1+l_2)/2}}{e^{-\alpha l_1} + e^{-\alpha l_2} - 2e^{-\alpha(l_1+l_2)/2}} = \left(\frac{1+e^{\alpha\Delta l/2}}{1-e^{\alpha\Delta l/2}}\right)^2 \approx \left(\frac{4}{\alpha\Delta l}\right)^2 \tag{4}$$

where $\Delta l = l_1 - l_2$. This approximation holds when $\alpha \Delta l$ is small. Therefore, the interferometric method allows the absorption coefficient $\alpha$ to be determined directly from the value of ER measured from the device without the need for determining the absolute input and output optical power.

The silicon photonic Mach-Zehnder interferometer devices used in the experiment were fabricated on a commercial silicon-on-insulator (SOI) substrate (Unibond, Soitec) using electron beam lithography and inductively coupled plasma etching. The thickness of the top silicon layer and the buried oxide layer (BOX) are 115 nm and 3 μm, respectively. The width of the waveguide is 500 nm, so only the fundamental quasi-TE mode can be guided. An important step for graphene integration is to planarize the photonic substrate with a cladding layer of annealed spin-on hydrogen silsesquioxane (HSQ)[28]. The planarization process flattens the surface of the substrate to prevent graphene from fracturing at the edges of the waveguides. In order to control the spacing between the graphene and the waveguide ($h$ in Fig 2b), this cladding layer was etched in buffered oxide etchant (BOE) to reduce its thickness to the desired value. The actual thickness of this cladding layer was further determined with a surface profilometer after opening a window using photolithography and wet etching.

Single layer graphene grown by chemical vapor deposition (CVD) on a copper foil was transferred to the photonic substrate using the method described by Li et al[11]. The PMMA layer that was used to assist the transfer was subsequently dissolved in acetone. Photolithography and oxygen plasma etching were used to pattern the graphene (Fig. 2) into patches of desired length on the silicon waveguide. To determine the Fermi level position and the carrier concentration in the transferred CVD graphene, a back-gated graphene field-effect transistor structure was fabricated on an n-Si/SiO$_2$ wafer. A typical drain current vs. gate voltage characteristic of such a device is shown in Fig. 2c. The result shows that typical transferred graphene layers are p-type doped with a background concentration of $\sim 4 \times 10^{12}$ cm$^{-2}$, which corresponds to a value of $E_F \sim -0.23$ eV. The strong p-type doping is a result of adsorbed moisture, and can be reduced by baking the sample in vacuum as shown in the Fig. 2c. However, since at the wavelengths ($\sim 1.55$ μm, $\hbar\omega = 0.8$ eV) utilized in this experiment, the condition of interband absorption $\hbar\omega/2 > |E_F|$ is satisfied, no vacuum bake was performed and the optical absorption measurements were conducted on the as-transferred samples. The Raman spectrum of a typical transferred graphene layer is also shown in Fig. 2d, confirming that the graphene is single layer.

We designed and fabricated unbalanced MZI devices with a fixed path length difference of $\Delta L = 100$ μm between the two arms, as shown in Fig. 3a. The graphene layer was subsequently transferred and patterned into rectangular patches on the waveguides of the interferometer arms. Fig. 3b shows a scanning electron microscope (SEM) image of a device. The area covered by graphene appears darker than the uncovered areas because of higher conductivity and consequently better discharge of electrons in graphene than in the oxide cladding. A laser source and a photodetector are coupled to the devices through integrated grating couplers (Fig. 3a) and a fiber array. The assumptions that the loss in the silicon waveguide is negligible and the graphene/waveguide hybrid in the two arms have the same $\alpha$ can be confirmed by measuring and comparing devices with conditions of $l_1 = l_2 = 0$ and $l_1 = l_2 \gg 0$, respectively. As shown in Fig. 3c, both devices shows similarly high extinction ratio of 35 dB, which indicates that the difference of $\alpha$ values in the two arms have an upper-bond value of 4%. In the device with $l_1 = l_2 = 70\,\mu\text{m}$, the higher insertion loss is apparently due to the high absorption loss in the graphene/waveguide hybrid, but the high extinction ratio also confirms the uniformity of the transferred graphene. Additionally, to confirm that the measurement is not affected by the possible residue of the removed PMMA layer, control samples were fabricated and tested by using the same transfer and PMMA removal processes but using a copper foil substrate without graphene grown on it. No discernible change in the extinction ratio of the transmission spectra was observed before and after the processes, indicating that residual PMMA is transparent and does not induce additional loss.

Next, we use a cut-back method and the above-described extinction ratio measurement to precisely determine the linear absorption coefficient in graphene. A large number (40) of devices were fabricated on each substrate and measured to accommodate the variation of the conditions of the transferred graphene. HSQ cladding layers were deposited and etched to a thickness, $h$, of 35 and 95 nm on two substrates. Graphene was then patterned into rectangular patches with an initial length of 150 μm on the waveguides in both interferometer arms (i.e. $l_1=l_2=150$ μm). Then, the length of graphene in the two arms was gradually reduced by photolithography and etching so that the differential length $\Delta l = l_1 - l_2$ was changed in multiple steps. In each step, the transmission spectra of all 40 MZI devices were measured to determine the ER value. Fig. 4a shows representative measured transmission spectra with varying $\Delta l$. As expected from equation (4), the extinction ratio in the interference fringes gradually decreases with the increasing $\Delta l$. The corresponding differential attenuation value $A = e^{-\alpha \Delta l}$ can be determined from the value of ER. To reveal the variation in all devices, histograms of the measurement results for different $\Delta l$ are shown in Fig. 4b. The variation in the measured values can be attributed to variability in the layer transfer process, including fluctuations in the flatness and continuity of the graphene. Since the distribution appears to be very close to a normal distribution, the average value along and the statistical error are plotted versus $\Delta l$ in the semi-logarithmic plot in Fig. 4c. The result shows an exponential dependence of the attenuation on $\Delta l$, yielding linear absorption coefficient $\alpha$ of 0.106 (1±3.3%) dB/μm and 0.0463 (1±4%) dB/μm for spacing layer thickness ($h$) of 35±8 and 95±6 nm, respectively. From equation (2) and the evanescent nature of the field outside the waveguide, the value of $\alpha$ is expected to be exponentially dependent on the distance between the graphene layer and the waveguide $h$. The result is plotted in Fig. 4d along with the theoretical value of $\alpha$ calculated using equation (2) and finite element simulation. An excellent agreement is obtained between the experimental results and the theory. The small discrepancy is attributed to the uncertainty of the actual values of $h$ due to non-ideal planarization. Extrapolating to zero distance ($h$=0) yields a maximal absorption coefficient of ~0.2 dB/μm. This value is comparable to the 0.1 dB/μm reported in ref. 12, where the discrepancy could be attributable to the difference in the waveguide geometry.

Our measurements confirm that, when integrated on a waveguide, graphene has a high absorption coefficient that can be determined from the universal optical conductivity as expressed by equation (2). These results have profound significance for the design of waveguide-coupled graphene optoelectronic devices. For instance, the zero-distance absorption coefficient of 0.2 dB/μm means that a graphene/waveguide structures needs only to be 15 μm long to obtain 3 dB attenuation and 50 μm to obtain 10 dB attenuation. To enhance this absorption further, as suggested by equation (2), the waveguide geometry and the mode profile can be engineered to increase the electrical field in the graphene, for instance, by utilizing a slot waveguide geometry. This high absorption coefficient indicates that high-efficiency, high-speed, broadband photodetectors and modulators can be achieved using the graphene-on-waveguide configuration with a small footprint. Our recent theoretical calculations have shown that a graphene-on-graphene modulator configuration can achieve bandwidth of more than 100 GHz at near-infrared (1.55 μm) and 30 GHz at mid-infrared (3.5 μm) wavelengths[29]. Thus, graphene integrated optoelectronic devices are very promising for high-capacity and low-power optical communication and interconnection applications using a wide optical spectral range from visible to mid-infrared.

**Acknowledgements**

SJK and YA would like to acknowledge the support of NSF and NRI under NSF Grant No. ECCS-1124831. ML acknowledges the faculty start-up fund provided by the College of Science and Engineering at the University of Minnesota. Parts of this work were carried out in the University of Minnesota Nanofabrication Center which receives partial support from NSF through NNIN program, and the Characterization Facility which is a member of the NSF-funded Materials Research Facilities Network via the MRSEC program.

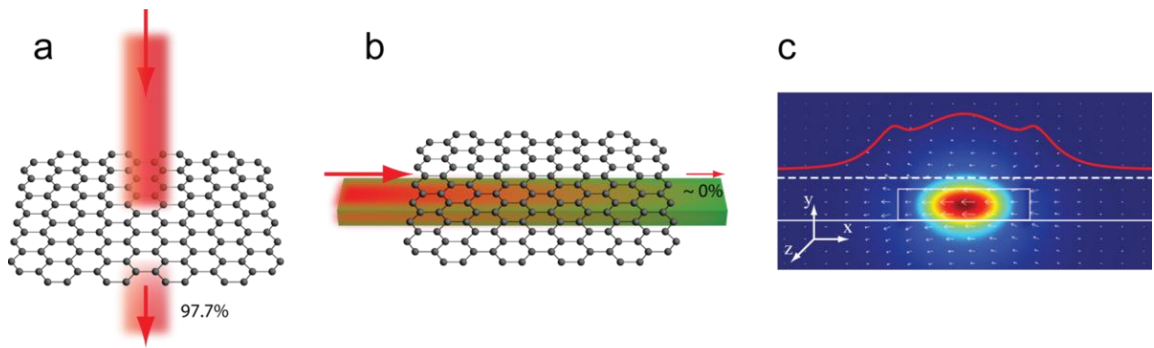

**Figure 1** a) For normal incident light, graphene has a universal absorption coefficient of 2.3%, but short interaction length. b) By integrating graphene on a waveguide, the light-graphene interaction length is only determined by the length of the device and so complete absorption can be achieved. c) Finite-element simulation result of the fundamental TE mode in a silicon waveguide with a layer of graphene (white dashed line) on top. The color map shows the optical intensity and the white arrows indicate the electric field. The red curve shows the profile of resistive dissipation $\langle Q_s(x) \rangle$ in graphene across its width as discussed in the text.

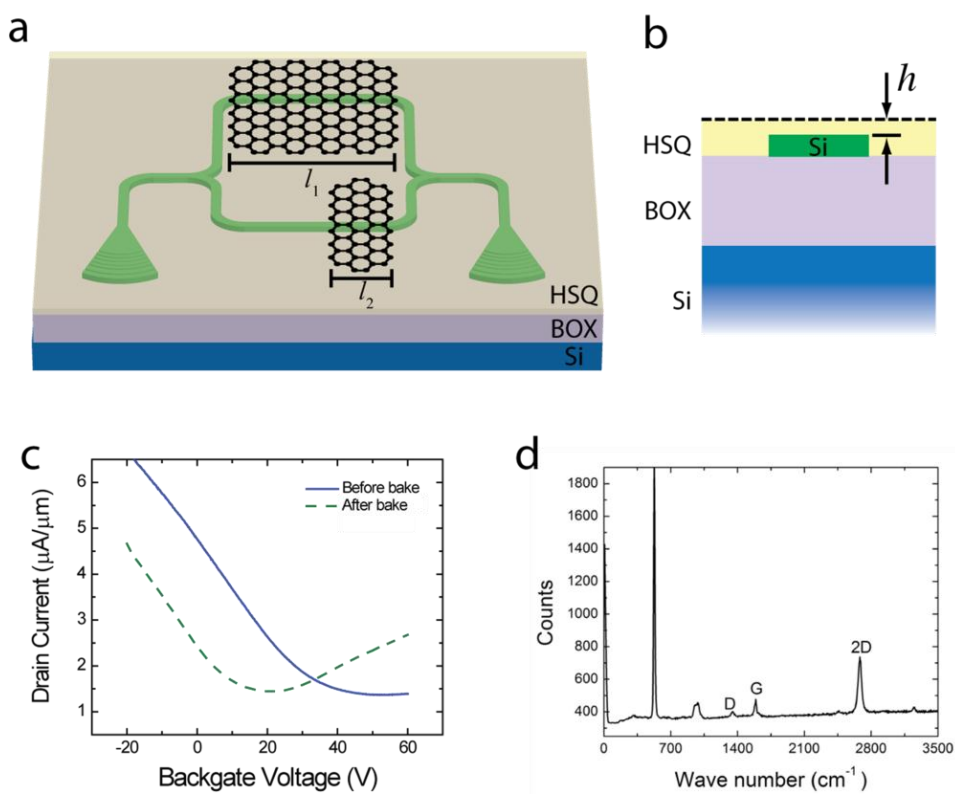

**Figure 2** a) An on-chip Mach-Zehnder interferometer made of silicon waveguides to precisely measure the absorption in graphene. The two interferometer arms are covered by graphene patterned to have different lengths. b) Cross-sectional view of the graphene/waveguide structure. Graphene is coated at a distance of $h$ above the waveguide with a layer of annealed HSQ cladding in between. c) Typical drain current versus backgate voltage curve measured in a field-effect transistor device made of transferred CVD graphene. The results show that typical transferred graphene layers are p-type doped with a background concentration of ~ $4 \times 10^{12}$ cm$^{-2}$, corresponding to a value of $E_F$ of –0.23 eV.  d) Raman spectrum of typical CVD graphene showing that the transferred graphene film is single layer.

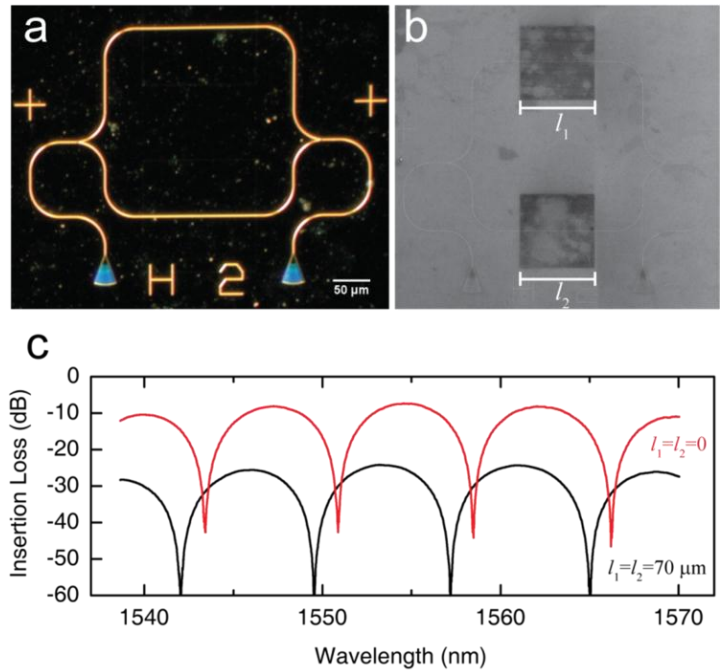

**Figure 3** a) Dark-field optical microscopy image of the silicon photonic Mach-Zehnder interferometer (MZI) device used in the experiment. b) Scanning electron microscopy image of the device after transfer and patterning of the graphene. The graphene covered area appears darker because of better discharging of the electrons, compared to the surrounding $SiO_2$ regions. c) Transmission spectra of two MZI devices. The first one (red) has no graphene coverage and the second one (black) has 70-µm-long graphene covering the waveguides in both arms of the interferometer. The equally high extinction ratio of 35 dB indicates that the absorption coefficient of graphene is similar in both arms and the loss in the silicon waveguide is negligible in comparison.

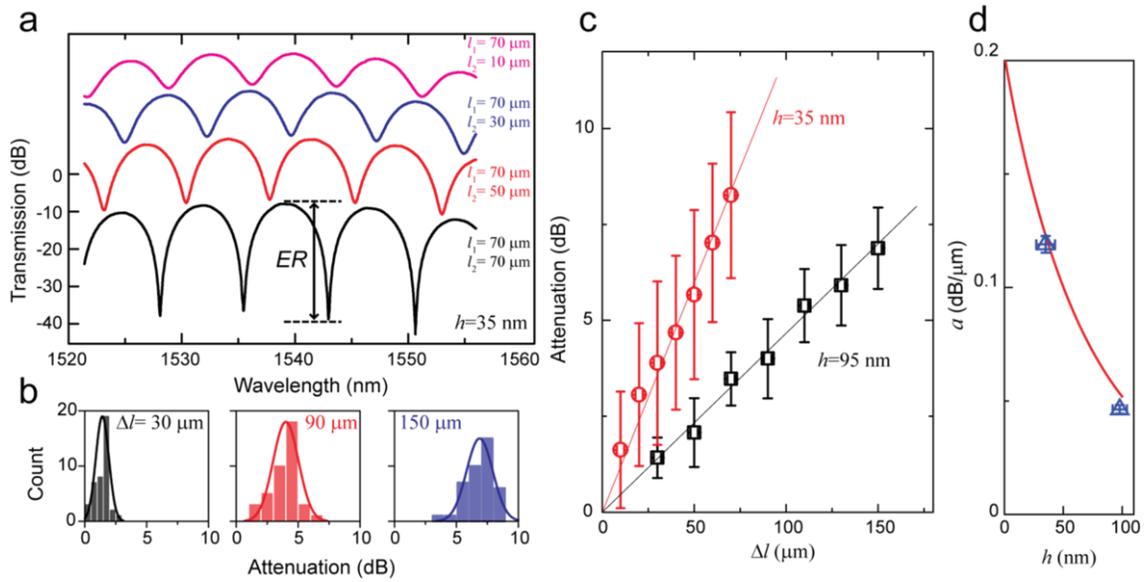

**Figure 4** a) Typical transmission spectra measured from devices with varying differential length $\Delta l$ of graphene-covered waveguide in the two arms of the MZI. A gradually decreasing extinction ratio is observed when the differential length $\Delta l$ is increased. The top three spectra (red, blue and cyan) are offset in the vertical direction for clarity. b) Histograms of linear attenuation $A = e^{-\alpha \Delta l}$ for different $\Delta l$ measured from 40 devices. c) Linear attenuation versus differential length for two values of HSQ spacing layer thickness, $h$. An exponential dependence is observed with the linear absorption coefficient $\alpha$ measured to be 0.106 $\times(1 \pm 3.3\%)$ dB/μm and $0.0463 \times (1 \pm 4\%)$ dB/μm for $h = 35(\pm 8)$ and $95(\pm 6)$ nm, respectively. The error bars are attributed to the variation of the conditions of the transferred graphene layer. d) Measured linear absorption coefficient $\alpha$ (blue symbols) versus $h$, showing a good agreement with the theoretical result (red line).